# Neutron Studies of the Iron-based Family of High $T_C$ Magnetic Superconductors

Jeffrey W. Lynn
NIST Center for Neutron Research, National Institute of Standards and Technology, Gaithersburg, MD 20899-6102

Pengcheng Dai
Department of Physics and Astronomy, University of Tennessee, Knoxville, TN 37996-1200 and Neutron Scattering Science Division, Oak Ridge National Laboratory, Oak Ridge, TN 37831-6393

ABSTRACT

We review neutron scattering investigations of the crystal structures, magnetic structures, and spin dynamics of the iron-based $R$Fe(As,P)O ($R$=La, Ce, Pr, Nd), (Ba,Sr,Ca)Fe$_2$As$_2$, and Fe$_{1+x}$(Te-Se) systems. On cooling from room temperature all the undoped materials exhibit universal behavior, where a tetragonal-to-orthorhombic/monoclinic structural transition occurs, below which the systems become antiferromagnets. For the first two classes of materials the magnetic structure within the *a-b* plane consists of chains of parallel Fe spins that are coupled antiferromagnetically in the orthogonal direction, with an ordered moment typically less than one Bohr magneton. Hence these are itinerant electron magnets, with a spin structure that is consistent with Fermi-surface nesting and a very energetic spin wave bandwidth ~0.2 eV. With doping, the structural and magnetic transitions are suppressed in favor of superconductivity, with superconducting transition temperatures up to ≈55 K. Magnetic correlations are observed in the superconducting regime, with a magnetic resonance that follows the superconducting order parameter just like the cuprates. The rare-earth moments order antiferromagnetically at low T like 'conventional' magnetic-superconductors, while the Ce crystal field linewidths are affected when superconductivity sets in. The application of pressure in CaFe$_2$As$_2$ transforms the system from a magnetically ordered orthorhombic material to a 'collapsed' non-magnetic tetragonal system. Tetragonal Fe$_{1+x}$Te transforms to a low T monoclinic structure at small x that changes to orthorhombic at larger x, which is accompanied by a crossover from commensurate to incommensurate magnetic order. Se doping suppresses the magnetic order, while incommensurate magnetic correlations are observed in the superconducting regime.

1. **Introduction**

The nature of the magnetic order and spin fluctuations in superconductors has had a rich and interesting history, and has been a topic of special interest ever since the parent materials of the high $T_C$ cuprates were found to be antiferromagnetic Mott insulators that exhibit huge exchange energies within the Cu-O planes [1]. These energetic spin correlations persist into the superconducting regime, often developing a 'magnetic resonance' whose energy scales with $T_C$ and whose intensity exhibits a superconducting order-parameter-like behavior [1]. The newly discovered iron oxypnictide superconductors possess a number of similarities to the cuprates, which naturally has led to strong parallels being drawn between the two classes of materials. There are, though, important differences as well. The basic properties are reviewed in detail elsewhere in this volume [2-4], so we simply highlight a few aspects that are key to understanding the magnetic properties exhibited by these materials. The iron-based systems are layered like the cuprates, although they are not nearly as two-dimensional in character (an important advantage for applications). The parent (undoped) materials exhibit long range collinear antiferromagnetic order, but are metallic rather than Mott insulators. Indeed all five iron $d$-bands are partially occupied and cross the Fermi surface, clearly classifying these materials as itinerant electron in character. Thus a multi-orbital theoretical description is necessary rather than the single orbital approach for the cuprates. The magnetic energies are very large, with a spin wave bandwidth ~0.2 eV. Moreover, in the superconducting regime, a 'magnetic resonance' excitation has been observed, just like for many of the cuprates. Here we review the neutron studies of the structure and magnetic transitions of the undoped materials, and how these progress with doping into the superconducting regime. We also discuss the spin dynamics that have been investigated, and the magnetic resonance that has been observed in the superconducting state. We note that all the properties discussed here have been corroborated by a variety of techniques such as resistivity, specific heat, magnetization, x-ray diffraction, Mössbauer spectroscopy, muon spin rotation measurements, and NMR, as described elsewhere in this review volume.

2. **Crystal and Magnetic Structures**

There are four different classes of iron-based superconductors typified by LaFeAsO (1:1:1:1), $SrFe_2As_2$ (1:2:2), LiFeAs (1:1:1), and $Fe_{1+x}$(Te-Se) (1:1). The (known) crystal structures at room temperature are all tetragonal [5-9], and are shown in Fig. 1. The important common aspect is that the $Fe^{2+}$ ions form square-planar sheets, where the direct iron-iron interactions render the $d$-electrons metallic in nature. The LiFeAs system is superconducting and does not order magnetically nor does the structure distort at low temperatures [7,8], as is the case for the related LaFePO systems [10]. The basic crystallographic information for these two systems is presented in Table 1, and we will not discuss them any further in this review. All the other undoped systems undergo a subtle structural distortion below room temperature that breaks the tetragonal symmetry. This transition is thought to be magnetically driven to relieve the magnetic frustration [3,4], and indeed long range magnetic order develops in the distorted state for all these materials. The structural phase transition temperatures are given in Table 2, along with the iron magnetic ordering temperatures, magnetic structures, and ordered moments in the ground state. Doping reduces and eventually completely suppresses these transitions as superconductivity develops.

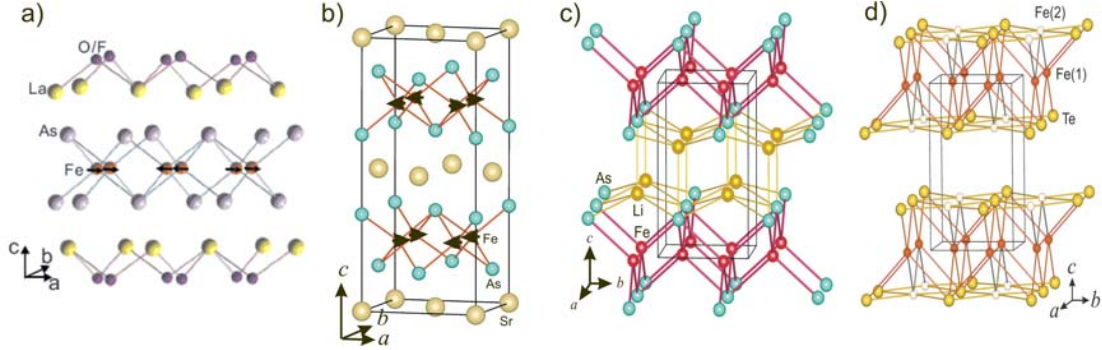

Fig. 1. Crystal structures for the four classes of superconductors: a) LaFeAsO, b) SrFe$_2$As$_2$, c) LiFeAs, and d) Fe$_{1+x}$Te. Below the tetragonal to orthorhombic structural transition the undoped materials can order magnetically, and the commensurate magnetic structures are also indicated in a) and b), where the spins are parallel along the orthorhombic *b* axis, antiparallel along the *a* axis, and with the spin direction along *a*. LiFeAs does not distort, while the magnetic structure for Fe$_{1+x}$Te is discussed below.

Table 1. Basic crystal structure of LaFeAsO and related materials. Above the structural transition the symmetry is tetragonal (*P4/nmm*), with Fe and O at special positions [2b, (3/4,1/4,1/2) and O 2b, (3/4,1/4,0), respectively], La [2c, (1/4,1/4,0.1417)], and As [2c, (1/4,1/4,0.6507)]. The internal coordinates are quite similar for the other 1:1:1:1 materials. At low T the structure is orthorhombic (*Cmma*), described by an *a-b* plane rotated by ≈45° and the lattice parameters unequal and multiplied by $\approx \sqrt{2}$. The special positions for the Fe and O can be generalized by employing the *P112/n* monoclinic space group which allows these atoms to shift along the *c*-axis. The crystal structure of SrFe$_2$As$_2$ is *I4/mmm*, with the Sr and Fe at the 2a (0,0,0) and 4d (1/2,0,1/4) special positions, respectively. The As occupies the 4e site, (0,0,0.3541). Below the structural distortion the 1:2:2 structure is orthorhombic *Fmmm*. LiFeAs is tetragonal *P4/nmm* at all T, Fe (2b, (3/4,1/4,1/2)), As (2c, (1/4,1/4,0.2635)), Li (2c, (1/4,1/4,0.8459). Fe$_{1+x}$(Te,Se) is tetragonal *P4/nmm* at elevated temperatures, Te (2a, (1/4,1/4,0.2829)) and two iron sites, Fe(1) (2b, (3/4,1/4,0)) and the partially occupied Fe(2) site, (2a, (1/4,1/4,0.7350)). Superconducting compositions remain tetragonal, while the non-superconducting ones distort at low T to monoclinic *P2$_1$/m* for smaller x, or orthorhombic for larger x.

| System | a (Å) | b(Å) | C(Å) | Ref. |
|---|---|---|---|---|
| LaOFeAs (175 K) | 4.0301 | ≡a | 8.7368 | [11-15] |
| (4 K) | 5.7099 | 5.6820 | 8.7265 | |
| CeOFeAs (175 K) | 3.9959 | ≡a | 8.6522 | [16] |
| (30 K) | 5.6626 | 5.6327 | 8.6382 | |
| PrOFeAs (175 K) | 3.977 | ≡a | 8.6057 | [19,20] |
| (5 K) | 5.6374 | 5.6063 | 8.5966 | |
| NdOFeAs (175 K) | 3.9611 | ≡a | 8.5724 | [17,18] |
| (0.3 K) | 5.6159 | 5.5870 | 8.5570 | |
| CaFe$_2$As$_2$ (175 K) | 3.912 | ≡a | 11.667 | [24-26] |
| | 5.542 | 5.465 | 11.645 | |
| SrFe$_2$As$_2$ (300 K) | 3.920 | ≡a | 12.40 | [21-23] |
| (150 K) | 5.5695 | 5.512 | 12.298 | |
| BaFe$_2$As$_2$ (175 K) | 3.9570 | ≡a | 12.9685 | [27-29] |
| (5 K) | 5.61587 | 5.57125 | 12.9428 | |
| LiFeAs (215 K) | 3.7914 | ≡a | 6.3639 | [7,8] |
| Fe$_{1.068}$Te (80 K) | 3.81234 | ≡a | 6.2517 | [9,30-33] |
| (5 K) | 3.83435 | 3.78407 | 6.2571 | |
| | | | β=89.212° | |

## 2.1. LaFeAsO and SrFe$_2$As$_2$ type systems

Neutron diffraction measurements have been carried out on the undoped La [11-15], Ce [16], Nd [17,18], and PrFeAsO [19,20] (1:1:1:1) materials, as well as the Sr- [21-23], Ca- [24-26], and BaFe$_2$As$_2$ [27-29] (1:2:2) systems. The crystal structure consists of single Fe-As layers that are separated by a single layer of (for example) LaO or Ba, respectively. They are all tetragonal at room temperature as already indicated, and undergo an orthorhombic distortion at lower temperatures. For the 1:2:2 materials the structural transition is clearly first-order in nature, and is directly accompanied by antiferromagnetic order. For the 1:1:1:1 systems the structural component of the ordering also appears to be first order, but the antiferromagnetic order generally develops at a lower temperature and appears to be second order.

The basic crystallographic information for the undoped materials of the four classes of systems is given in Table 1. The lattice parameters are given for the high temperature tetragonal phase and for the low temperature distorted structure. For atoms that are not at special positions, the internal coordinates quoted in the caption are for the prototype systems, namely LaFeAsO, SrFe$_2$As$_2$, LiFeAs, and Fe$_{1.068}$Te, in the tetragonal (higher T) phase. These internal coordinates are representative and do not vary substantially between high and low temperature, or for different cations. If more detailed crystallographic information is needed then the references should be consulted.

The crystallographic distortion breaks the tetragonal symmetry, and the materials become orthorhombic. In this crystallographic description, the Fe and O ions remain at special positions. The refinements from the initial study [11] suggested that the O ion might be shifted, which can be described by the monoclinic *P2/c* space group, allowing the Fe and O ions to have a displacement along the *c*-axis. Most powder diffraction studies do not indicate that this additional degree of freedom is necessary, and higher precision single crystal diffraction studies would be desirable to determine which description is best. But this is a rather subtle feature, and at the present level of precision either description can be employed.

The structural distortion is thought to be driven by the magnetic interactions, as the lower symmetry relieves the magnetic frustration and allows the system to order [3,4]. The observed magnetic structure within the *a-b* plane is identical for both classes of materials and consists of chains of Fe spins that are parallel to each other along the (short) *b*-axis of the distorted tetragonal cell, while along the longer *a*-axis the spins are coupled antiferromagnetically, with the spin direction along the *a*-axis as shown in Fig. 2. Note that this type of magnetic structure is forbidden in tetragonal symmetry. The observed spin structure is consistent with Fermi-surface nesting, although the calculated ordered moment based on first-principles theory is much larger than observed [3,4]. This is another indication that these metals are itinerant electron magnets, which simply means that the electrons which are unpaired and magnetically active occupy energy bands that cross the Fermi energy. Along the *c*-axis the nearest-neighbor spins can be either antiparallel as for the La and Nd 1:1:1:1 systems, or parallel like Ce and Pr. For the three 1:2:2 materials nearest neighbors along the *c*-axis are antiparallel. The magnetic configurations are all simple commensurate structures.

For the 1:2:2 materials sizable single crystals are available that enable more detailed investigations of the structural and magnetic phase transitions, such as shown in Fig. 3. The intensity of the tetragonal peak for SrFe$_2$As$_2$ is followed as the structural transition is traversed, and the sudden change in intensity shows that the structural transition is clearly abrupt (first order). In the distorted phase, the T dependence of the (1,0,1) magnetic peak shows that the magnetic order develops at the same temperature as the structural distortion. On first inspection the magnetic order parameter looks continuous (second order), but it is actually truncated just at $T_N$, and the lack of any significant critical scattering also reveals the first order nature of the magnetic transition.

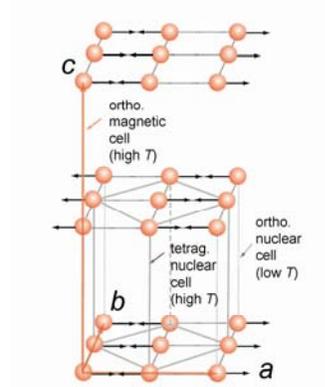

Fig. 2. Magnetic structure for the iron spins in the 1:1:1:1 and 1:2:2 systems. The in-plane spin configuration and spin direction are identical for all these materials, where the spins are parallel along the orthorhombic *b* axis, antiparallel along the *a* axis, and with the spin direction along *a*. Along the more weakly coupled *c*-axis the arrangement can be either parallel (ferro) or antiparallel (antiferro). All the structures are simple commensurate magnetic structures.

Table 2. Structural phase transitions for undoped materials, together with the ordering temperatures, spin configuration, and ordered moment for the iron spins. As a function of doping, all studies so far have found that both the structural and iron magnetic phase transitions decrease with increasing doping level. We note that some samples have been found to be unintentionally doped, and low transition temperatures for the nominally undoped materials have been reported. Here we list the higher T observations. The magnetic rare earth ions in the 1:1:1:1 materials order at low temperatures in commensurate magnetic structures, and the ordering temperature, magnetic moment and ordering wave vector are also given.

| Material | $T_S$ (K) | $T_N$(Fe) (K) | $\mu_{Fe}$ ($\mu_B$) | $q_{Fe}$ | Spin direction | $T_N(R)$ (K) | $\mu_R$ ($\mu_B$) | $q_R$ | Spin direction | Ref |
|---|---|---|---|---|---|---|---|---|---|---|
| LaOFeAs | 155 | 137 | 0.36 | 101 | likely *a* | - | | | | [11-14] |
| CeOFeAs | 158 | 140 | 0.8 | 100 | *a* | 4.0 | 0.94 | 101 | *a,b,c* | [16] |
| PrOFeAs | 153 | 127 | 0.48 | 100 | *a* | 14 | 0.84 | 100 | *c* | [19] |
| NdOFeAs | 150 | 141 | 0.25 | 101 | likely *a* | 1.96 | 1.55 | 100 | *a,c* | [17,18] |
| CaFe$_2$As$_2$ | 173 | 173 | 0.80 | 101 | *a* | - | | | | [24-26] |
| SrFe$_2$As$_2$ | 220 | 220 | 0.94 | 101 | *a* | - | | | | [21-23] |
| BaFe$_2$As$_2$ | 142 | 143 | 0.87 | 101 | *a* | - | | | | [27] |
| Fe$_{1.068}$Te | 67 | 67 | 2.25 | 100 | *b* | - | | | | [32] |

For both classes of materials where it has been determined, the easy axis (spin direction) is along *a* as indicated in Table 2. This determination rests on the ability to distinguish *a* from *b* for the magnetic peaks. The size of the ordered moment is typically one $\mu_B$ or considerably less, and recalling that the magnetic intensity is proportional to the square of the ordered moment, the moment direction has not yet been determined for some of the smaller-moment 1:1:1:1 materials because of the weak magnetic scattering in the powders. For the three 1:2:2 materials, it is straightforward to distinguish *a* from *b* on the available single crystals, and the spin configuration is the same as for the 1:1:1:1 systems, with the spins parallel along the shorter *b*-axis, antiparallel along the longer *a*-axis within the *a-b* plane, and with the spin direction along *a* as shown in Fig. 1. It is likely that all these materials have the spin direction along *a*.

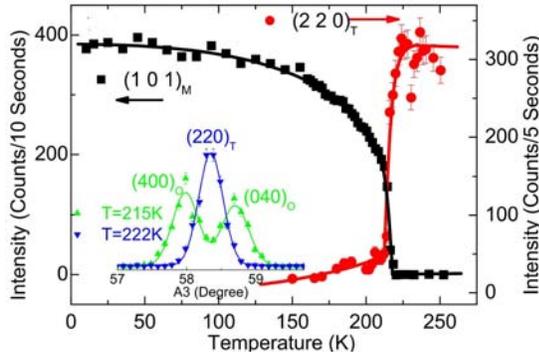

Fig. 3. Peak intensity of the tetragonal (220) peak in a single crystal of $SrFe_2As_2$ as a function of temperature (solid circles). The crystallographic distortion splits the peak into the (400) and (040) orthorhombic peaks, which causes the intensity in between to rapidly decrease. At the same temperature the development of the (101) magnetic peak signals the onset of long range magnetic order [21].

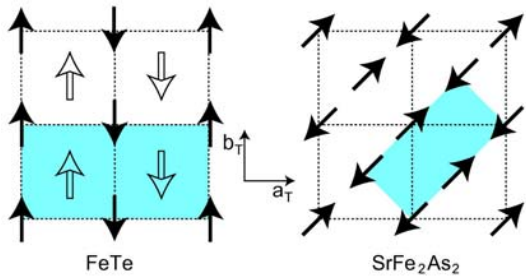

Fig. 4. Magnetic structure for $Fe_{1.068}Te$, which is commensurate with the underlying lattice. The magnetic configuration and spin direction differ from the 1:1:1:1 and 1:2:2 magnetic structures [32].

2.2. $Fe_{1+x}$(Te-Se) system

For the $Fe_{1+x}$(Te-Se) system [9, 30-33], crystallographically there are two iron sites, one of which is partially occupied, while the (Te, Se) site is fully occupied. Hence the composition should be indicated as $Fe_{1+x}$(Te,Se). This material is tetragonal at elevated temperatures. For the pure Te system the structure distorts into a monoclinic phase, where commensurate antiferromagnetism abruptly sets in at the same temperature. Thus the magnetic and structural transitions occur simultaneously in a first-order transition. In addition, the two inequivalent Fe sites are both active magnetically. The magnetic structure is shown in Fig. 4, and the nature of the distortion as well as the magnetic structure the system exhibits are different than the 1:1:1:1 and 1:2:2 systems. This contrasts with theoretical expectations based on a 1:1 stoichiometry, but the difference may be due to the additional iron site. Interestingly, at higher Fe content (for example $Fe_{1.141}Te$ [31]) the magnetic order becomes incommensurate, and the incommensurability wave vector is strongly dependent on the Fe content. With Se doping the structural distortion changes from monoclinic to orthorhombic while the magnetic order is suppressed in favor of superconductivity. However, incommensurate spin fluctuations survive into the superconducting regime [31]. The pure FeSe superconducting phase is stoichiometric (x=0) [33].

2.3. $CaFe_2As_2$ under pressure

The application of modest pressure (a few kbars) was found to cause $CaFe_2As_2$ to go superconducting [34]. Pressure dependent neutron diffraction measurements revealed a dramatic change in the crystal structure—a strongly first-order phase transition to a "collapsed" tetragonal phase [25,26]. By collapsed we mean that there is a huge decrease in the *c*- axis lattice parameter, by 10%, and an overall decrease in the volume of the unit cell by 5%; the *a-b* plane undergoes a smaller expansion. The region of superconductivity appeared to occur in this collapsed phase. However, the initial reports of superconductivity were carried out on a single crystal with a solid medium providing the pressure, before the large changes in the crystal structure were discovered. Because of the huge anisotropic change in the lattice as the system is transformed into the collapsed tetragonal phase, the pressure applied using a solid medium to produce the superconductivity is also hugely anisotropic. Subsequent measurements under hydrostatic pressure revealed that the superconducting phase was completely or nearly absent under these conditions [35]. The detailed origin of the superconductivity is now not resolved, but it appears to occur in a mixed-phase region. It would be interesting to investigate the superconductivity using epitaxially grown thin films where the appropriate stress can be applied.

One of the very interesting aspects of this collapsed phase is that first principles calculations using the observed crystal structure indicate that the Fe moment itself collapses [3,4,25,26]. Indeed, the neutron diffraction measurements do not find any evidence for magnetic order in the tetragonal collapsed phase [25,26], and inelastic scattering data do not find any evidence for spin correlations [36]. If the collapsed phase is actually superconducting, it's very likely that magnetic fluctuations cannot be the origin of the pairing. It will be interesting to see if spin correlations do exist when anisotropic pressure is applied, and in what crystallographic phase the superconductivity is present. These may be particularly difficult measurements for neutron scattering on single crystals, however, because the pressure has to be applied in the *a-b* plane.

2.4. Rare Earth Magnetic Ordering

For the 1:1:1:1 systems, the rare earth ordering has been studied for the Ce [16], Nd [18], and Pr [19,20] materials. They all order at low T like "conventional" magnetic superconductors; $T_N(Ce) = 4$ K, and $T_N(Nd) = 2$ K. It is interesting, however, that the ordering temperature for Pr is much higher than the other rare earth ions, $T_N(Pr) = 14$ K, much like what happens in the 1:2:3 cuprate superconductors where $T_N(Pr) = 17$ K [37]. The Pr spins order in the rather complicated magnetic structure as shown in Fig. 5, where trios of spins above and below the plane that contains the oxygen ions are coupled ferromagnetically, while adjacent trios align antiferromagnetically. The spin direction is simply along the *c*-axis, and adjacent planes of spins are identical along the *c*-axis so that the magnetic unit cell is the same as the nuclear unit cell. Then the ordering wave vector is [1,0,0] in the orthorhombic system. The Nd system exhibits the same type of spin configuration, but with components of the ordered moment along all three axes rather than just along *c*. The Ce magnetic structure, on the other hand, has moments primarily in the *a-b* plane, with a chain-like structure similar to the iron, but with adjacent chains with their spin direction approximately orthogonal rather than antiparallel. The Ce ions may also have a small component of the moment along the *c*-axis. The direction of the spins for nearest neighbors along the *c*-axis is also (approximately) orthogonal rather than parallel, so that the magnetic unit cell is doubled and the ordering wave vector is [1,0,1]. The rare earth magnetic ordering temperatures, ordered magnetic moments, and ordering wave vectors are given in Table 2.

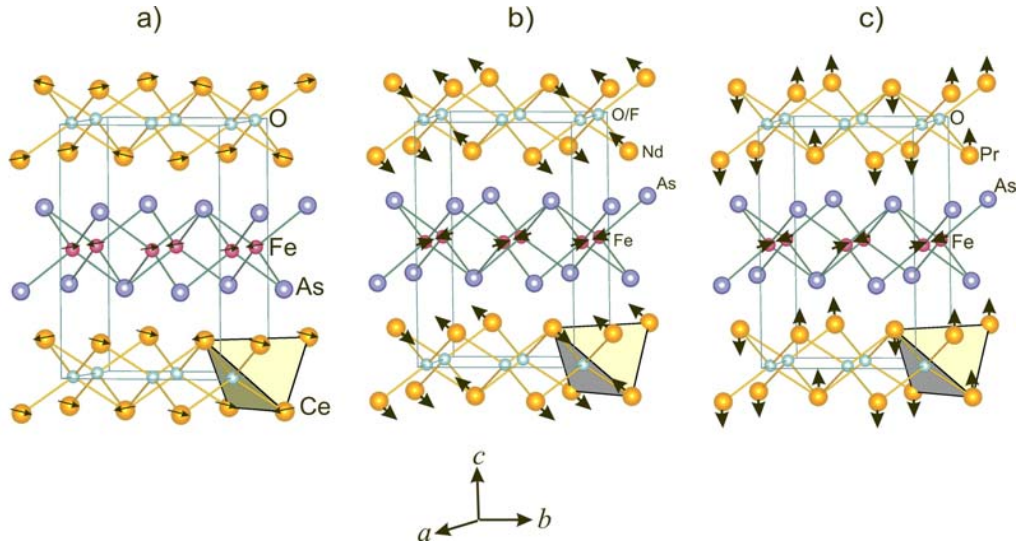

Fig. 5 Low temperature magnetic structures for the rare earth and iron moments in a) CeFeAsO, b) NdFeAsO, c) PrFeAsO. In each, the iron magnetic structure is assumed not to change when the rare earth moments order.

It should be noted that the rare earth and iron spins interact with each other, which can alter both magnetic structures and correlate the moments obtained in the refinements. In addition, there are a limited number of magnetic peaks with two relatively small moments to refine. Hence in the refinements the spin structure for the iron sublattice was assumed to be the same as found above the rare earth ordering, while the size of the moment was allowed to vary. In the case of the Ce and Nd systems this produced iron moments substantially larger than found above $T_N$ (rare earth), but it is not clear at this stage whether this is a real increase or an artifact of the limited magnetic data and concomitant assumptions. Data on larger samples/longer counting times might be helpful to clarify this issue, but it's final resolution may have to await the availability single crystals large enough for neutron diffraction (or resonant x-ray diffraction).

The rare earth ordering of the Ce has also been investigated systematically for the doped system. Interestingly, as the iron transition temperature approaches zero with increasing fluorine content (discussed below), the Ce moment rotates from the *a-b* plane to being along the *c*-axis. There is little effect on the transition temperature for the Ce order, which contrasts with the Pr system both as a function of F doping and oxygen depletion, where no Pr magnetic order is observed down to 5 K. Thus the Pr ordering temperature has decreased dramatically with doping.

2.5  Doping Dependence of the Structural and Magnetic Transitions

Initial studies revealed that the structural distortion and long range magnetic order were absent in the optimally doped LaFeAsO$_{1-x}$F$_x$ material [11], and this was found to be the case for all the 1:1:1:1 and 1:2:2 materials investigated to date [11,12,14,16,19,38,39]. The doping dependence of the structural and magnetic transitions has been investigated in detail for the La [14] and Ce [16] 1:1:1:1 materials, and the phase diagram for CeFeO$_{1-x}$F$_x$ is shown in Fig. 6. The structural and magnetic temperatures both decrease with increasing doping content, with the iron Néel temperature decreasing more rapidly. For the Ce system [16], it is apparent that the long range order for the iron vanishes before superconductivity appears. Therefore these two order parameters appear to fully compete with each other. For the La 1:1:1:1 system the transition as a function of doping may be first order or there could be coexistence [14], while for the Ba$_{1-x}$K$_x$Fe$_2$As$_2$ there is evidence of coexistence of antiferromagnetic order and superconductivity [39]. On the other hand, it is clear that the orthorhombic structural phase overlaps into the superconducting regime for these systems, so that the superconductivity can occur in both the tetragonal and orthorhombic structures.

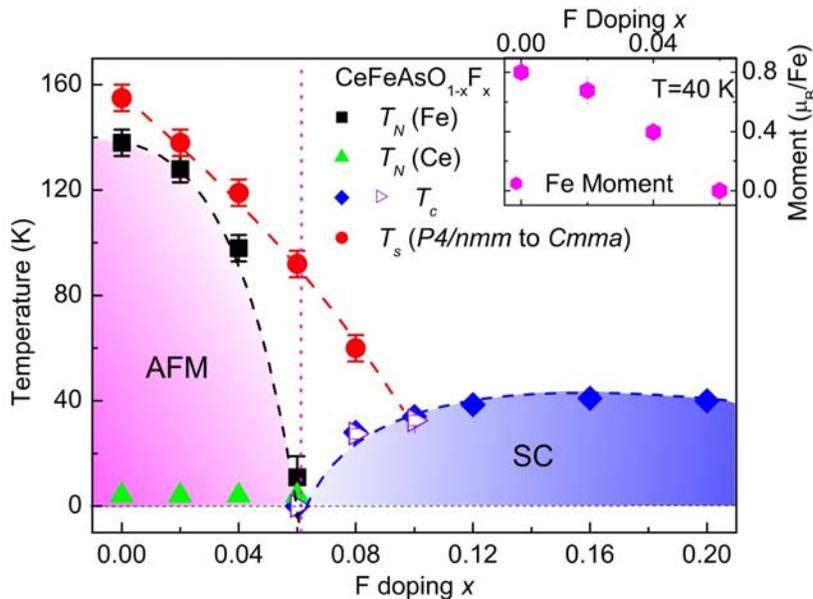

Fig. 6. Phase diagram for CeFeAsO$_{1-x}$F$_x$ as a function of fluorine doping. Both the structural and magnetic phase boundaries decrease with increasing x. The magnetic long range order is suppressed before superconductivity develops, while the superconductivity is able to develop in the orthorhombic as well as the tetragonal structure [16].

One trend that has become apparent in the crystallographic studies is a systematic decrease in the Fe–As/P–Fe bond angle for Fe-based superconductors with higher T$_C$ as shown in Fig. 7 [16], indicating that lattice effects play an important role in the superconductivity. Indeed the highest T$_C$ is obtained when the Fe–As/P–Fe angle reaches the ideal value of 109.47° for the perfect FeAs tetrahedron. This suggests that the most effective way to increase T$_C$ in Fe-based superconductors is to decrease the deviation of the Fe–As/P–Fe bond angle from the ideal Fe-As tetrahedron. It also explains in a pedagogical manner that we have reached the maximum T$_C \approx 55$ K for these single-layer iron arsenide materials.

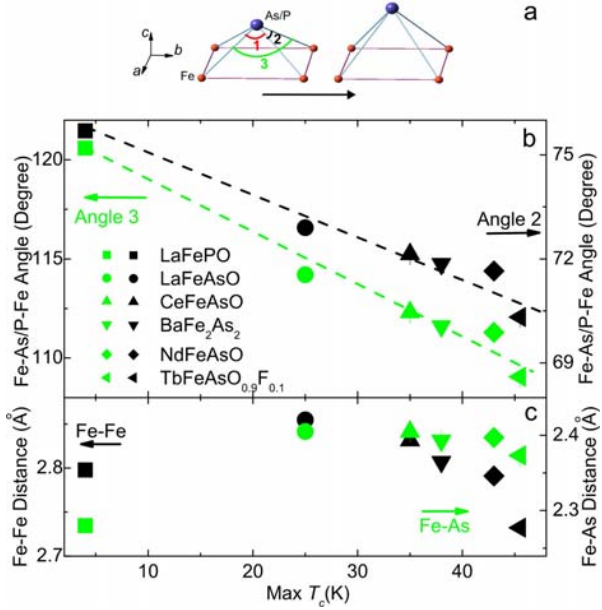

Fig. 7. The Fe-As(P)-Fe bond angle is found to vary systematically with $T_C$ for the Fe-based superconductors. a) Schematic illustration of what happens to the Fe-As-Fe tetrahedron as a function of $T_c$. b,c) Dependence of the maximum-$T_c$ on the Fe-As(P)-Fe angle and Fe-Fe/Fe-As(P) distance. The maximum $T_c$ is obtained when the Fe-As(P)-Fe bond angle reaches the ideal value of $109.47°$ for the perfect FeAs tetrahedron [16].

## 3. Inelastic Scattering Studies

For the cuprate superconductors, the undoped materials are Mott insulators where the Cu $S=1/2$ spins are two-dimensional in nature, with a spin wave bandwidth that is typically ~0.2-0.4 eV [1]. The undoped iron-based superconductors are also magnetically ordered, and the magnetic exchange interactions are only somewhat smaller than the cuprates. More interestingly, the spin correlations in both systems survive into the superconducting regime, with a magnetic resonance that is clearly linked to the superconducting order parameter.

3.1. Iron Spin Waves

Single crystals are available for the 1:2:2 materials (and for the 1:1 systems) that are not only large enough for neutron diffraction studies, but inelastic studies as well. For the undoped materials the spin wave dispersion relations have been measured for the $SrFe_2As_2$ [40], $CaFe_2As_2$ [41,42], and $BaFe_2As_2$ [43,44] systems, and the overall spin dynamics for the three systems are quite similar as shown in Table 3. These studies reveal that the spin dynamics are quite energetic, with a spin wave bandwidth ~0.2 eV as shown in Fig. 8 for $SrFe_2As_2$. This energy scale is comparable to the energy scale of the $S=1/2$ Cu spins in the Mott insulating cuprates. In contrast to the cuprates, though, there is significant spin wave dispersion along the *c*-axis, although the overall dispersion is still anisotropic. Hence these materials are not strictly two-dimensional like the cuprates, but are better described as anisotropic three-dimensional materials. All three materials also have a significant spin gap in the antiferromagnetic spin wave spectrum of the parent material, and the origin of this gap has not been established yet.

Table 3. Low temperature spin dynamics results, using Eq. (1), for the undoped 1:2:2 materials, obtained on powders (pwd) and single crystals (xtl). For the Ba velocities, the data were obtained just above the spin wave gap, and hence it is likely that the actual values of the velocities will be larger when more complete data are available.

| Material | $v_{a-b}$ (meV-Å) | $v_c$ (meV-Å) | $\frac{v_c}{v_{a-b}}$ | Band width (meV) | Gap (meV) | Reference |
|---|---|---|---|---|---|---|
| $SrFe_2As_2$ xtl | 560±110 | 280±56 | 0.5 | 170 | 6.5(2) | [40] |
| $CaFe_2As_2$ xtl | 420 ±70 | 270±100 | 0.4 | 200 | 6.9(2) | [41,42] |
| $BaFe_2As_2$ xtl | 280±150 | 57±7 | 0.2 | | 9.8(2) | [44] |
| $BaFe_2As_2$ pwd | | - | | 175 | 7.7(2) | [43] |

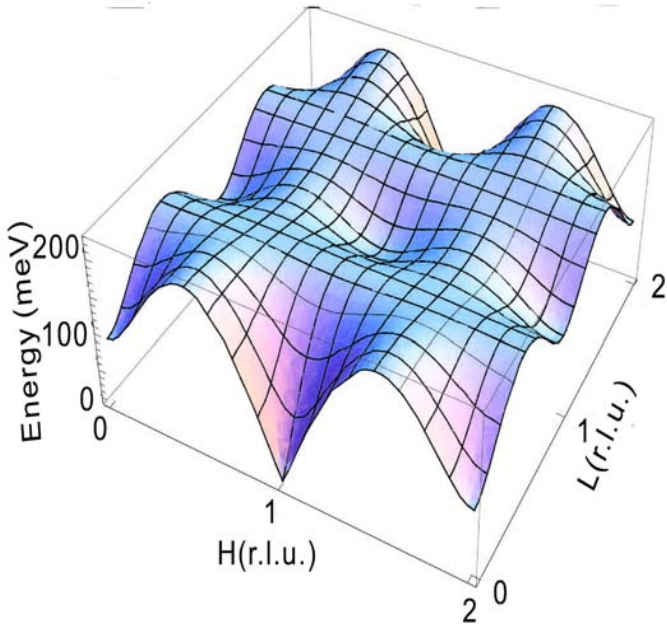

Fig. 8. Calculated spin wave dispersion relations for $SrFe_2As_2$ based on inelastic neutron scattering measurements taken in the low energy regime [40]. The bandwidth of the spin waves is ≈0.2 eV for all three materials.

These materials are itinerant electron systems and consequently the usual description of the spin excitations would be in terms of the dynamic susceptibility $\chi(\mathbf{q},\omega)$ of the electron system. However, because of the large overall bandwidth of the spin wave excitations the higher resolution measurements have been taken at relatively small wave vectors and a simple parameterization of the dispersion relations in terms of a Heisenberg model and linear spin wave theory can be used, employing the empirical relation

$$E(\vec{q}) = \sqrt{\Delta^2 + v_{ab}^2\left(q_x^2 + q_y^2\right) + v_c^2 q_c^2} \qquad (1)$$

where the energy $E(\mathbf{q})$ and the spin wave gap $\Delta$ are in meV, the $v_{ab}$ and $v_c$ are the in-plane and c-axis spin wave velocities in units of meV-Å, and the wave vectors are in Å$^{-1}$. The results for the three materials are given in Table 3, where we see that the basic description is quite similar for the three in terms of the spin wave dispersion and overall bandwidth of the spin waves. Obviously these first experimental results will be refined as the measurements are extended, but the basic energetics are now known. At high energies the spin waves are reported to be heavily damped [42], as can occur for itinerant electron systems where the collective excitations interact with the single-particle density of states. We would also expect that the overall spin dynamics for the 1:1:1:1 systems will be similar, but at the present time sizable single crystals are not available and little work has been carried out [15]. However, spin dynamics work is now ongoing for the Fe(Se-Te) materials.

3.2. Superconducting Regime

The initial search for magnetic correlations in the superconducting regime carried out on a small polycrystalline sample of $LaFeAsO_{0.87}F_{0.13}$ was unsuccessful [18]. A subsequent measurement on a very large polycrystalline sample of $Ba_{0.6}K_{0.4}Fe_2As_2$ observed magnetic correlations, and the development of a magnetic resonance in the superconducting state [45], and this work was quickly followed by single crystal measurements [46-48] as shown in Table 4. The observation of a magnetic resonance that is directly associated with the formation of the superconducting state, analogous to the magnetic resonance phenomena first discovered in the cuprates, makes it clear that magnetic fluctuations play an intimate role in the superconductivity of both classes of high $T_C$ superconductors.

Table 4. Inelastic neutron scattering results for the magnetic resonance $E_r$ that develops in the superconducting phase of the 1:2:2 materials.

| Material | $E_r$(meV) | $T_C$(K) | $E_r/k_BT_C$ | Reference |
|---|---|---|---|---|
| $Ba_{0.6}K_{0.4}Fe_2As_2$ pwd | 14 | 38 | 4.3 | [45] |
| $BaFe_{1.84}Co_{0.16}As_2$ xtl | 9.6(3) | 22 | 5.1 | [46] |
| $BaFe_{1.9}Ni_{0.1}As_2$ xtl | 9.1(4) | 20 | 5.3 | [47,48] |

There have not been any measurements of the dispersion of the rare earth magnetic excitations in any of these materials, but there have been detailed measurements of the energies and linewidths of the 18 meV Ce crystal field level in superconducting $CeFeAsO_{0.84}F_{0.16}$, with some interesting results [49]. Below the superconducting transition a substantial change in the energy and increase in the intrinsic linewidth of the level was observed. This can be understood as the coupling of superconducting electrons with the crystal field levels. When the superconducting gap opens an energy renormalization can be expected, along with a decrease in the linewidth for any excitation below the gap, because the electrons form a bound (Cooper) pair and there is not enough energy to break the pair. On the other hand, an increase in linewidth would be expected for excitations above the gap, due to a 'piling-up' of the available one-electron states. This energy renormalization and linewidth behavior have been observed for both phonons [50] and crystal field levels [51] in conventional electron-phonon superconductors. In the present case the 18 meV level should be above the gap, and hence an increase in the linewidth would be expected, as observed. A dramatic splitting of the crystal field levels has also been observed for the undoped system when the magnetic system orders.

## 4. Future Directions

This is a very young field as far as the superconductivity is concerned, but the pace of research has been extremely rapid. At this point the basic physics of the undoped (parent) 1:1:1:1 and 1:2:2 materials is fairly well established in terms of the structure, magnetic order, and spin dynamics, and in terms of how these overall properties compare with expectations based on first-principles theoretical calculations. Of course, there are many issues still to be quantified and resolved. The spin waves need to be measured to higher energies for all three of the 1:2:2 materials, and determine if the damping is from single-particle (Stoner) excitations as might occur for itinerant electron systems, while little work on the spin dynamics of the 1:1:1:1 materials is available yet. Much progress also has been made on the doping dependence of the properties, in that it is clear that the structural and magnetic phase transitions are reduced and eventually disappear as the superconductivity develops. It appears that the magnetic order does not overlap with the superconductivity in some systems, but it apparently does in others. This coexistence could be macroscopic in origin, or it could be intrinsic, and this issue will only be resolved when the question of inhomogeneity is fully addressed; the question is whether these two order parameters are mutually exclusive. Magnetic correlations do persist into the superconducting regime, and the elucidation of the magnetic fluctuation spectrum in the superconducting regime is one of the most important areas to explore. The magnetic fluctuations are the present frontrunner among possible pairing mechanisms, as is the case for the cuprates; both systems are highly correlated electron materials—the cuprates more so. However, the pairing mechanism is by no means settled for either system; one only has to recall that the pairing mechanism in the cuprates is still elusive after more than two decades of intensive research. There may be some surprises for the community, and perhaps the iron-based superconductors will provide the key to understanding both classes of materials.

The discovery of high temperature superconductivity in these iron-based materials has focused the attention of the condensed matter physics community on these new superconductors. Although the superconductivity is new, they belong to an enormous class of systems with a wide variety of properties. This will enable the physical properties to be tailored both to investigate the fundamental properties of these systems as well as for applications. This flexibility provides a vast potential that will stimulate the field for the foreseeable future.

**Acknowledgments**

The authors would like to express their sincere gratitude to all of their collaborators, as listed in the references, who have shared the research excitement of this new family of superconductors. P.D. is supported by the US Department of Energy, Division of Materials Science, Basic Energy Sciences, through DOE DE-FG02-05ER46202; and by the US National Science Foundation through DMR-0756568. This work is also supported in part by the US Department of Energy, Division of Scientific User Facilities, Basic Energy Sciences.